\documentclass[12pt,preprint]{aastex}

\newcommand{\kepler}{{\it Kepler}}

\shorttitle{\kepler{} Mission Signal Detections}
\shortauthors{Tenenbaum et al.}

\begin{document}

\title{Detection of Potential Transit Signals in the First Twelve Quarters
of \kepler{} Mission Data}

\author{Peter Tenenbaum, Jon M. Jenkins, Shawn Seader, Christopher J. Burke, 
Jessie L. Christiansen, Jason F. Rowe, 
Douglas A. Caldwell, Bruce D. Clarke, Jie Li, Elisa V. Quintana, 
Jeffrey C. Smith, 
Susan E. Thompson, and Joseph D. Twicken}
\affil{SETI Institute/NASA Ames Research Center, Moffett Field, CA 94305, USA}
\email{peter.tenenbaum@nasa.gov}
\author{William J. Borucki, Natalie M. Batalha, Miles T. Cote, Michael R. Haas, Roger C. Hunter, and Dwight T. Sanderfer}
\affil{NASA Ames Research Center, Moffett Field, CA 94305, USA}
\author{Forrest R. Girouard, Jennifer R. Hall, Khadeejah Ibrahim, Todd C. Klaus, Sean D. McCauliff, Christopher K. Middour, 
Anima Sabale, Akm Kamal Uddin, and Bill Wohler}
\affil{Orbital Sciences Corporation/NASA Ames Research Center, Moffett Field, CA 94305, USA}
\and
\author{Thomas Barclay and Martin Still}
\affil{BAER Institute/NASA Ames Research Center, Moffett Field, CA 94305, USA}

\begin{abstract}
We present the results of a search for potential transit signals in
the first three years of photometry data acquired by the \kepler{} Mission.
The targets of the search include  
112,321 targets which were observed over the full interval and an additional
79,992 targets which were observed for a subset of the full interval.  
From this set of targets we find
a total of 11,087 targets which contain at least one signal which meets the Kepler detection criteria: those criteria
are periodicity of the signal, an acceptable signal-to-noise ratio, and three tests which reject
false positives.  Each target containing at least one detected signal is then searched repeatedly for additional
signals, which represent multi-planet systems of transiting planets.  When targets with multiple detections are
considered, a total of 18,406 potential transiting planet signals are found in the \kepler{} Mission dataset.
The detected signals are dominated by events with relatively low signal-to-noise
ratios and by events with relatively short periods.  The distribution of estimated
transit depths appears to peak in the range between 20 and 30 parts per million, with
a few detections down to fewer than 10 parts per million.  The detections exhibit
signal-to-noise ratios from 7.1 $\sigma$, which is the lower cut-off for detections,
to over 10,000 $\sigma$, and periods ranging from 0.5 days, which is the shortest period searched,
to 525 days, which is the upper limit of achievable periods
given the length of the data set and the requirement that all detections include at least 3 transits.
The detected signals are compared to a set of known transit events in the
\kepler{} field of view, many of which were identified by alternative methods; 
the comparison shows that the current search recovery rate for targets with known 
transit events is  98.3\%.
\end{abstract}



\section{Introduction}

We have previously reported \citep{pt2012} on the results of searching the
first 218 days of \kepler{} Mission \citep{wjb2010} data for potential signals indicative of transiting
planets.  In the intervening time, there have been two developments in the
search for potential exoplanets in the \kepler{} dataset.  First, the algorithms used in
the \kepler{} analysis pipeline have undergone dramatic improvements.  Second, the
data available for searching has expanded from 218 days to 1050.5 days.  
This massive increase in data volume makes possible searches for 
exoplanets with much longer orbital periods, as well as searches for extremely small
exoplanets with relatively short period orbits.  In this study we report on the results of searching
the current set of \kepler{} observations with the upgraded analysis pipeline.  This study
can be considered as an update of the previous report \citep{pt2012}.

\subsection{\kepler{} Science Data}

The operational parameters of the \kepler{} Mission have been extensively reported \citep{science-ops}.  
In brief:  the \kepler{} spacecraft is in an Earth-trailing
heliocentric orbit of 372 day period.  Its single instrument, the \kepler{} photometer,
points almost constantly at a 115 square degree region of the
sky centered on $\alpha = 19^{\rm h}22^{\rm m}40^{\rm s}, \delta = +44.5\degr$.  During
science operations, photometric data is taken in 29.4 minute integrations, known within
\kepler{} as ``long cadences'' (as distinguished from ``short cadences,'' which are 1/30 of
a ``long cadence'' and are collected for a small subset of targets).  
In order to 
maintain the correct orientation of the solar panels and thermal radiator, the spacecraft
rotates about the photometer boresight axis by $90\degr$ approximately every 93 days, the interval at a given
orientation being referred to as a ``quarter.''  A consequence of this rotation is that each
target star is observed each year on 4 different readout channels on the focal plane.  
Science acquisition is interrupted for monthly downlinking of
pixel data, maneuvering from one quarter's attitude to the next, reaction wheel desaturation
(one 29.4 minute sample is lost for this purpose approximately every 3 days), and a
variety of spacecraft anomalies.  

The data acquisition period for this analysis begins at 2009 May 12 00:00:00 UTC, ends
at 2012 March 28 12:47:26 UTC, and contains 51,412 sample intervals of 29.4 minutes.
Of these, 47,588 intervals are dedicated to science data acquisition, the balance of 3,824
intervals being consumed by the interruptions listed above.  During this period the 
spacecraft performed 11 axial rotations, resulting in 12 quarters worth of data.

\begin{figure}
%\epsscale{.80}
\plotone{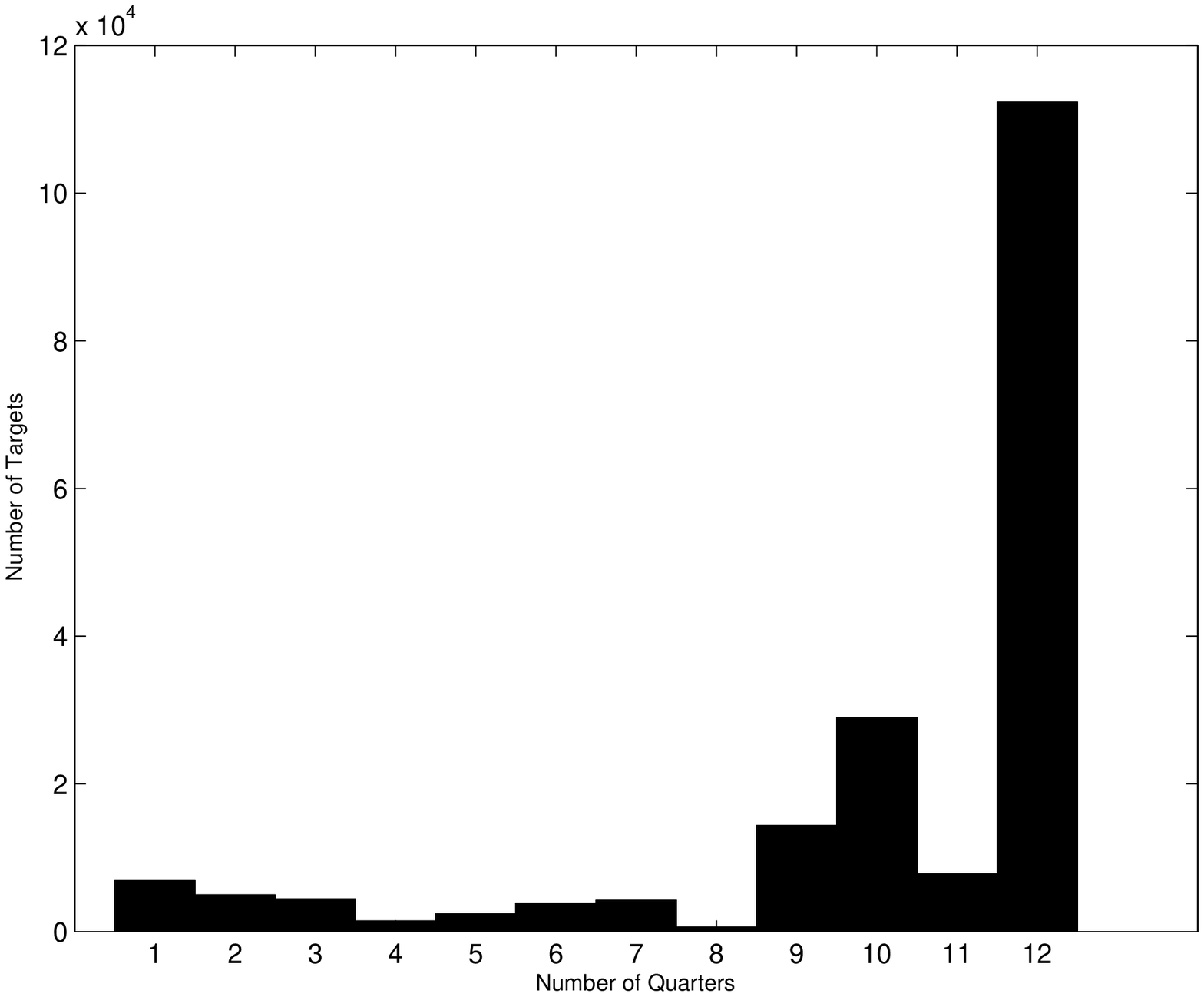}
\caption{Histogram of number of quarters of observation for all targets.  The significant number of targets
observed for 10 quarters out of 12 is primarily due to an onboard electronics failure which prevents
readout from 4 out of the 84 CCD modules on the focal plane, resulting in a ``blind spot'' which rotates through the
field of view as \kepler{} rotates about its axis.
\label{f1}}
\end{figure}

A total of 192,313 targets were observed by \kepler{} during the 12 quarters of data acquisition, 
and were subsequently searched for indications of transiting planets.
Of those, 112,321
were observed in all 12 quarters; the balance of 79,992 were observed only in a subset
of quarters.  Figure \ref{f1} shows the distribution of targets according to the number of
quarters observed.
Observation of a target in a subset of quarters can occur for any of three reasons.
The most significant cause of limited observation is an onboard electronics failure which occurred
on 2010 January 23, one month into quarter 4:  this failure resulted in the subsequent loss 
of all data from 4 of the 84 CCD readouts on the focal plane (specifically, the 4 CCD readouts in 
Module 3).  
Due to the quarterly rotation
of the spacecraft, this failure produced a ``blind spot'' in the \kepler{} field of view which 
moves relative to the target stars, causing a large number of targets to be visible only
75\% of the time.  Any target which falls onto Module 3
was only observed in 10 out of 12 quarters. 
The 28,965 stars which were observed for 10 quarters as shown 
in Figure \ref{f1} are mainly due to this effect.
A second limitation on the number of quarters for which a target is observed is that 
the process of target selection and prioritization has evolved over the life of the
\kepler{} Mission; targets which are added or removed subsequent to Quarter 1 will not be 
observed during all quarters.  
Additionally, a fraction of \kepler{}'s observing capacity is reserved for use by the \kepler{} Guest Observer
(GO) and Asteroseismic Science Consortium (KASC) programs; the targets observed in these programs
are frequently updated, resulting in a number of targets observed for relatively short intervals.
Finally, due to small asymmetries
in the construction of the focal plane, a small number of targets cannot be observed in all 
spacecraft orientations:  in some quarters these targets are imaged onto one or another CCD detector,
while in some quarters the target images fall between the detectors.  In total, 28,826 targets were
observed in 8 or fewer quarters; 43,339 targets were observed for 9 or 10 quarters; and 7,819
targets were observed for 11 quarters.

In addition to the aforementioned 192,313 targets which were searched for planets, a total of 
2,123 known eclipsing binaries which were observed but not searched for
transiting planet signatures.  This was done for operational reasons.  The \kepler{} processing
pipeline has limited capacity to identify circumbinary planets because their transit
signatures are generally neither periodic nor of constant duration.  
However, the eclipses of an eclipsing binary system mimic planetary transits with sufficient fidelity
to be identified by TPS as potential signals of transiting planets.  
These known eclipsing binaries were removed
to reduce the computational and human burden which would otherwise have been imposed by
their false-positive detections.

\subsection{Pre-Search Processing}

The processing of pixel data from the \kepler{} spacecraft, prior to the search for transiting planet
signatures, is summarized elsewhere \citep{pipeline}.  The processing step which has seen the most dramatic
change is Pre-Search Data Conditioning (PDC).  The purpose of PDC is to remove
variations in the flux time series which are generated by changes in the spacecraft
environment or other systematic effects.  The original PDC algorithm determined the systematics by 
performing a robust least-squares fit of assorted spacecraft engineering variables to each flux time series,
and then subtracting the systematics thus determined to yield a conditioned flux time series \citep{jdt2010b}.
While such an approach is guaranteed to reduce the bulk RMS variation of each target's flux, it can also distort
the true stellar variations and can even add variability on timescales of interest for planet
searches.  Both of these unwanted side effects are driven by the same source: the least-squares fit is removing variability
which is {\it coincidentally} correlated with some engineering variable, but not {\it causally} related.  

This unwanted behavior is corrected by applying a Bayesian approach to constrain the fitted amplitudes of
systematic error terms which are then removed from the light curves.  This process allows the algorithm to 
deduce ``reasonable'' values for the correlation of each identified systematic to the light curves, and thus to reject
correlations which are wildly out of family.  Additionally, the ensemble of target star data across a large number
of stars is used to empirically identify systematic trends in the light curves, rather than relying upon the available
spacecraft engineering data.  The algorithm is fully described elsewhere \citep{js2012,martin2012}.

In addition to the corrections described above, the current PDC algorithm identifies and corrects the signature
of a cosmic ray related artifact known as a Sudden Pixel Sensitivity Dropout (SPSD).  An SPSD occurs when 
a cosmic ray produces a step reduction in the quantum efficiency of a pixel; the reduction is typically of order
one percent, and the quantum efficiency partially recovers, typically over a period of hours to days.  Because an SPSD
bears a superficial resemblance to a transit signature (at least to a computer), efficient removal of SPSDs without
inadvertent removal of actual transits is a crucial step in data conditioning for \kepler{}.  Unlike environmental
signatures, SPSDs are completely uncorrelated from one target star to another, and thus are removed from the
data via a separate algorithm within PDC.  

\section{Transiting Planet Search}

The Transiting Planet Search (TPS) algorithm
is described in some detail in \citet{jmj2002} and \citet{jmj2010}, 
as well as \citet{pt2012}.  The improvements in the algorithm since \citet{pt2012} are summarized below.

\subsection{Edge Detrending of Contiguous Blocks of Flight Data}

The algorithm which was previously used to remove trends at the ends of single-quarter data segments
was replaced with an algorithm which performs a robust
fit of the form:
\begin{equation}\label{eq1}
y = P_1 \exp( -x/P_2 ) + P_3x + P_4 + P_5 \exp[ (x-1)/P_6 ],
\end{equation}
where $y$ is the median-corrected flux, $x$ is the sample time normalized to a range from 0 to 1, and $P_1$ 
through $P_6$ are the parameters of the fit.  In words, Equation \ref{eq1} fits a line plus two
exponential edge trends, one at the leading edge of the data region and one at the trailing region, with both
the amplitude and the time constant of the exponentials as fit parameters.  The form in Equation \ref{eq1} was found
to match the actual edge trends as well as the constrained polynomial fit which had previously been used.  The
advantages of the reformulated edge-trend removal are: a reduced number of assumptions and/or configuration
parameters for the fit; use of the full data segment for the entire fit; robust fitting; and the fact that the new fit does not
under any circumstances introduce a polynomial ``wave'' into the data segment in an attempt to correct the edges (i.e.,
over fitting).  Additionally, whereas in the past the edge detrending was applied only to full quarters of data, in the 
current implementation it is applied at any time when there was an interruption of data acquisition to change the
spacecraft orientation.  This was done to mitigate the thermal transients which occur when the spacecraft attitude is
changed.  Attitude change incidents include all data downlink intervals, plus any transitions into or out of safe mode.

\subsection{Detection and Vetoing of Potential Signals}

The first step in detection of potential signals is described in Section 2 of \citet{pt2012}:  a wavelet-based, adaptive matched
filter is utilized to search for periodic reductions in flux occurring against the non-white, non-stationary background of 
stellar variability.  The significance of such a reduction is known as its Multiple Event Statistic.
The Multiple Event Statistic is computed across a two-dimensional grid of signal period and epoch of first transit, and across
14 trial transit pulse durations; the maximum Multiple Event Statistic from this set is captured, along with the combination of period, epoch, 
and transit pulse duration (henceforth "signal timing") which generated it. 
A threshold is then applied to the maximum Multiple Event Statistic to reject targets which are unlikely to contain a true
transiting planet signature.  The threshold value represents a balance between rejecting true positives in the event of an 
excessively high threshold versus accepting false positives in the event of an excessively low threshold.  This balance was
extensively explored prior to \kepler{} launch \citet{jmj2002}.  Based on these studies, a threshold of 7.1 $\sigma$ was adopted for the 
Multiple Event Statistic.  At this threshold, the probability of detecting an Earth-sized planet which produces 4 transits of a 12th magnitude
Sun-like star is approximately 80\%; the expected false alarm probability from statistical fluctuations is at the level of 1 false alarm
per 600,000 target-years of observations, which translates to 1 false alarm detection during the entirety of the nominal \kepler{} mission.
Target stars for which the maximum Multiple Event Statistic falls below the specified detection threshold of 7.1 $\sigma$ are
rejected from further analysis.  The requirement that the maximum Multiple Event Statistic exceed
7.1 $\sigma$ removes from further consideration 76,668 targets, leaving 115,645 with at least one potential transit signal which 
lies above this threshold.  

The principal weakness of the Multiple Event Statistic calculation is that it cannot discriminate between a true train of transit
events (which have uniform depth, duration, and shape to within the precision limits of the instrument) and a chance combination
of dissimilar events which coincidentally occur within a flux time series.  As an example, consider a flux time series for which the
combined differential photometric precision (CDPP) for transit detection is 50 parts per million (PPM) at all times \citep{jessie}.  
If the flux time
series contains 4 uniformly-spaced transits of with uniform depths 250 PPM, the resulting Multiple Event Statistic for that period and epoch will be
10 $\sigma$, and will be reported as an above-threshold event by the Multiple Event Statistic calculation.  
On the other hand, if the 4 transits are uniformly spaced but do not have uniform depth -- for example, if the depths of the 4 transits
are 20 PPM, 30 PPM, 50 PPM, and 900 PPM, respectively -- the Multiple Event Statistic for this combination of events will also be
10 $\sigma$, and will also be reported as an above-threshold event by the Multiple Event Statistic calculation.
While the former scenario might be the signature of a transiting planet, the latter clearly is not.
Thus, a Multiple Event Statistic which is above the detection threshold is a necessary but not sufficient condition for identifying a 
potential transiting planet signature.  More generally, while the matched filter approach is optimal with respect to rejecting the null
hypothesis, it is insufficient for discrimination between competing alternate models.  For this reason,
once a Multiple Event Statistic above the detection threshold is identified, the event thus detected 
is subjected to a series of tests which are designed to discriminate between potential transit
signatures and heterogeneous combinations of unrelated events.  These tests accept the former while vetoing the latter.

\subsubsection{Robust Statistic Veto of False Positive Detections}

As described above, detections due to transiting planets and false alarm detections can be separated from one another by the requirement that
the transits are periodic, of equal duration, and uniform depth.  The Multiple Event Statistic calculation strongly enforces the requirement of periodicity
and weakly enforces the requirement of uniform duration, but as described above does not enforce the requirement of uniform depth.  The
Robust Statistic veto is complementary to the Multiple Event Statistic test in that it tests each detection for uniformity of transit depth.  This is accomplished
by constructing a model flux time series with transits, in which the transits are represented by square pulses which have the epoch, period, and duration
dictated by the signal timing of the Multiple Event Statistic.  The model flux time series is fitted to the data, with the transit depth being the only
free parameter in the fit.  In order to eliminate the effect of stellar variations, both the flux time series and the model transit
pulse train are whitened, as described in \citet{jmj2010}.  A robust fit is utilized in order to reduce the influence of out-of-family samples in the flux
values which participate in the fit.  The Robust Statistic, which is the signal-to-noise ratio estimated from the fit, is then used to reject false positives.  For a 
more complete description of the Robust Statistic see Appendix \ref{robustStat}.  Specifically, a large value of the Robust Statistic indicates a detection in which 
the transits are reasonably uniform in depth and duration, which is characteristic of true transit signatures; a small value indicates that the Multiple Event Statistic 
has been formed from a combination of heterogeneous transit-like events with unequal depths, which is characteristic of false positives.  The Robust Statistic threshold was selected using the results of an earlier TPS exercise with fewer quarters of data: the Robust Statistics for 
targets known to have true-positive transiting planets were compared to those for other stars with Multiple Event Statistics above the threshold of 7.1 $\sigma$.  
The value of 6.4 $\sigma$ caused 98\% of the former to be accepted, while rejecting 66\% of the latter.  Increasing the threshold above this value 
caused an unacceptable number of known true-positive detections to be rejected.  Note that this method of tuning the Robust Statistic threshold 
implicitly assumes that the latter set of detections is so dominated by false alarms, and contains so few true positive detections, that it is safe to
treat the set as being entirely false alarms; given that over 100,000 target stars produced Multiple Event Statistics in excess of 7.1 $\sigma$, this seems
a safe assumption.  In the current TPS run, a threshold of 6.4 $\sigma$ for the Robust Statistic rejects 79,030 targets, leaving 36,614 targets which require further scrutiny.  

\subsubsection{$\chi^2$ Veto of False Positive Detections}

In the second test used for vetoing of false positives, the signal which produced the Multiple Event Statistic is decomposed is two different ways, 
namely, first into its wavelet 
scale contributions for each transit and second into its temporal contributions.  For a true transit event with the period, epoch, transit duration, and 
Multiple Event Statistic of the detected signal, and assuming uniform transit depths, it is possible to compute the expected values in each of these decompositions.
As shown briefly in Appendix \ref{chisquare}, and more thoroughly in
\citet{seader2012}, the expected component values for each transit are compared to observed values in the construction of two functions, each
of which is expected to be distributed according to a $\chi^2$ distribution.  These functions are then combined with the Multiple
Event Statistic of the potential signal, as shown in Appendix \ref{chisquare}.  By requiring that the values of the two resulting discriminators,
$X_{(1)}$ and $X_{(2)}$, both exceed 7.0, we veto an additional 25,506 targets, yielding 11,108 targets which contain potential transiting
planet signatures.  Note that these thresholds were tuned empirically in a manner identical to that used to tune the Robust Statistic, described above.  An event
 which has passed all four tests -- Multiple Event Statistic, Robust Statistic, and $\chi^2$ discriminators -- is
referred to as a Threshold Crossing Event (TCE).

\subsection{Iterative Rejection of False Positives and Re-Searching of the Flux Time Series}

Prior versions of TPS suffered from a significant design weakness:  in cases in which the strongest transit-like feature
was vetoed, 
the search of that target would terminate.  In this way a strong but low-quality transit-like signal could inadvertently 
mask a weaker but higher-quality event.  This flaw is addressed in the current version of TPS:  in the event that an
apparent transit is vetoed, TPS goes on to search for additional transit signatures in the same light curve.  Because the
search of additional periods and epochs can potentially be extremely time-consuming, for operational purposes it is
necessary to limit the number of iterations of searching which are permitted for a given target and a given trial transit
pulse duration.  At present the limit is set to 1000 iterations of re-searching.  In the analysis reported here, approximately
three quarters of all TCEs 
occurred on the first iteration of the search, with the balance TCEs detected on subsequent iterations.  
The largest number
of iterations required to detect a TCE was 404.  

\subsection{Removal of Non-Periodic Transit-Like Features}

\begin{figure}
%\epsscale{.80}
\plotone{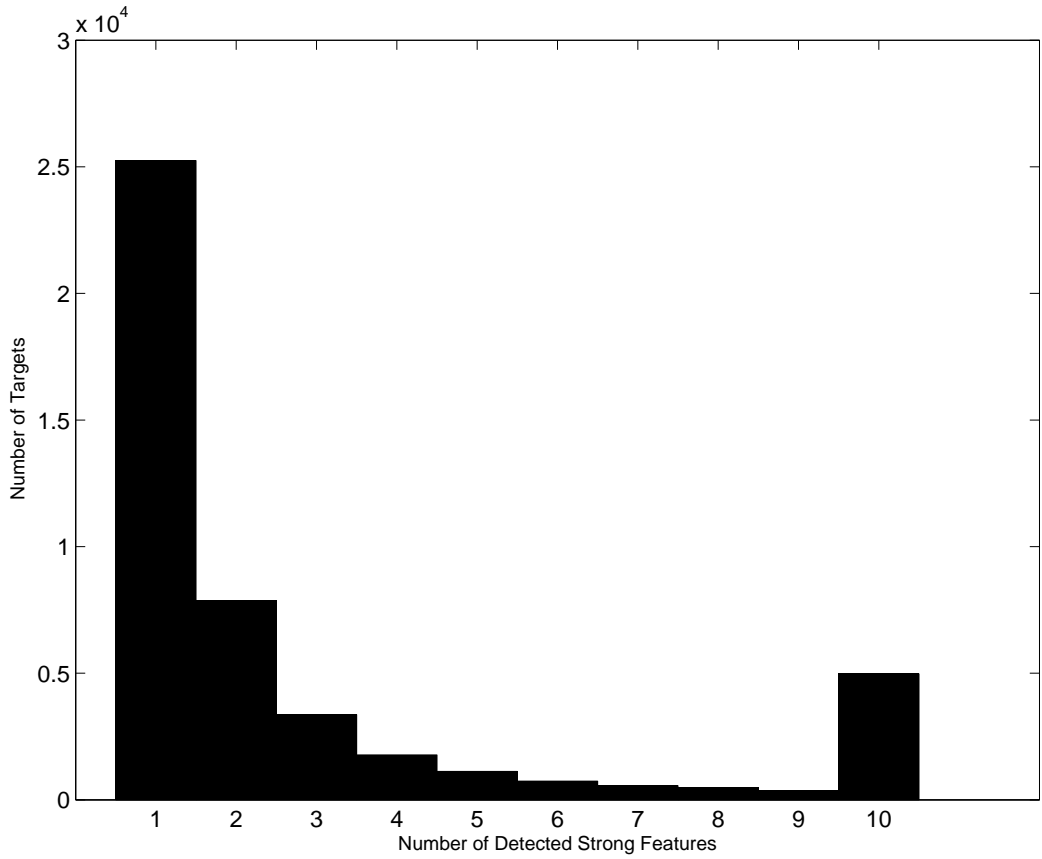}
\caption{Distribution of the number of strong features in each flux time series, as defined in the text.  The
final bin includes overflows:  there are a total of 327 targets with 10 features and 4,650 with more than
10 features.
\label{f2}}
\end{figure}

The benefits of the multiple iterations of search, described above, can only be fully exploited in the absence of relatively
strong non-astrophysical single events.  Such strong events will cause the Multiple Event Statistic to exceed the 7.1 $\sigma$
threshold for large numbers of possible periods:  folding a single strong event with a small number of weak events will 
produce a large Multiple Event Statistic, and there are an extremely large number of period-epoch combinations which will
result in such a folding.  If this happens, the 1000 iterations of searching can easily be exhausted in the process of eliminating
a fraction of the spurious Multiple Event Statistics caused by a single strong event.  Such an outcome can be
avoided if these strong events are identified and removed prior to folding, but such removals are obviously dangerous:
without prior knowledge, a feature in the data which is identified as a non-astrophysical event, and removed, could actually
be a strong transit.  For this reason, any event removal must be used sparingly.  TPS addresses this issue in two ways.  First,
a minimum number of transits is required for an event to be accepted, since the probability of such chance combinations
yielding a Multiple Event Statistic over threshold decreases as the number of events folded together increases.  At present,
the threshold number of transits is 3.  Second, the current version of TPS is permitted to 
remove one, and only one, single event, and only in the case in which the first iteration of planet searching produces a strongest
event which exceeds the Multiple Event Statistic threshold of 7.1 $\sigma$ but which is then vetoed by RS, $X_{(1)}$, or
$X_{(2)}$.  In such a case the strongest single event in the time series is removed, if and only if the strongest single event
has an amplitude which is greater than the Multiple Event Statistic threshold multiplied by the square root of the minimum number of
transits (7.1 $\sigma\times\sqrt{3}$, or 12.3 $\sigma$ for the current parameter choices).  Out of the 11,108 TCEs, 2,193 are
found on light curves which 
have had such a feature removed.  Additionally, on each target the number of such identifiable features is counted and recorded, 
regardless of whether any such events are removed.  
Out of all 192,313 targets, the number which have at least one identifiable strong single
event is 46,481.  Figure \ref{f2} shows the distribution of the number of strong events for targets which have at least one such event.
Note that the distribution is strongly peaked towards small numbers of events, implying that it is worth considering the option of
using more aggressive removal of features in future TPS runs.

\subsection{Limitation on Allowable Transit Duty Cycles}

During development of the most recent version of TPS, it was observed that a substantial number of false positives were 
produced with a short period and a long trial transit pulse duration, which implied that applying a threshold to the ratio of 
trial transit pulse duration to period (henceforth known as the "transit duty cycle") would allow suppression of a large number 
of false positive detections.  The transit duty cycle for a central transit of the Sun by the Earth is approximately $7.4\times10^{-4}$, which
implies that for Earth-analogues the threshold could be set to an extremely low value; however, for a given star the 
transit duty cycle is inversely proportional to the semi-major axis of the transiting body's orbit, and thus setting a low threshold for the transit duty 
cycle will implicitly eliminate sensitivity to short-period planets.  For a solar-type star, the minimum TPS search period of 0.5 days would
lead to a transit duty cycle of 0.092 for a circular orbit; the TPS duty cycle should therefore be somewhat larger than this value in order to preserve 
sensitivity to 0.5 day orbits on larger stars and to allow some margin for eccentric orbits.  Given these considerations,
for the processing run reported on here we limited the transit duty cycle to values below 0.16 .  

\subsection{Detection of Multiple Planet Systems}

In \citet{hw2010}, the process for detection of multiple planet systems is described.  In brief, for each target star which yields
a valid detection as described above, a planet model is fit to the flux time series, using the period and epoch of the TCE as a
starting point for the fit; the transit signatures from the fitted planet model are removed from the flux time series; and the residual flux time series is then searched 
for additional TCEs.  The subsequent TCE search is performed using the same TPS algorithm as is used for the initial search.
When multiple planet detections are included, the total number of TCEs increases to 18,427.  

Following the detection and model fitting described above, an additional set of automated analyses are performed which 
allow astrophysical false positives, such as background eclipsing binaries, to be ruled out.  For the purposes of the discussion below,
we will consider only the TCEs for which the additional automated analyses were successfully completed:  this set includes
18,406 TCEs falling on 11,087 targets.  Of the 21 excluded targets, 19 are non-stellar ``super-aperture'' targets, for which the
automated post-detection analyses cannot be performed, while 2 are conventional \kepler{} targets for which the automated
post-detection analyses failed due to software errors.  
Each of the excluded targets produced a single TCE.

\section{Detected Signals of Potential Transiting Planets}

\begin{figure}
%\epsscale{.80}
\plotone{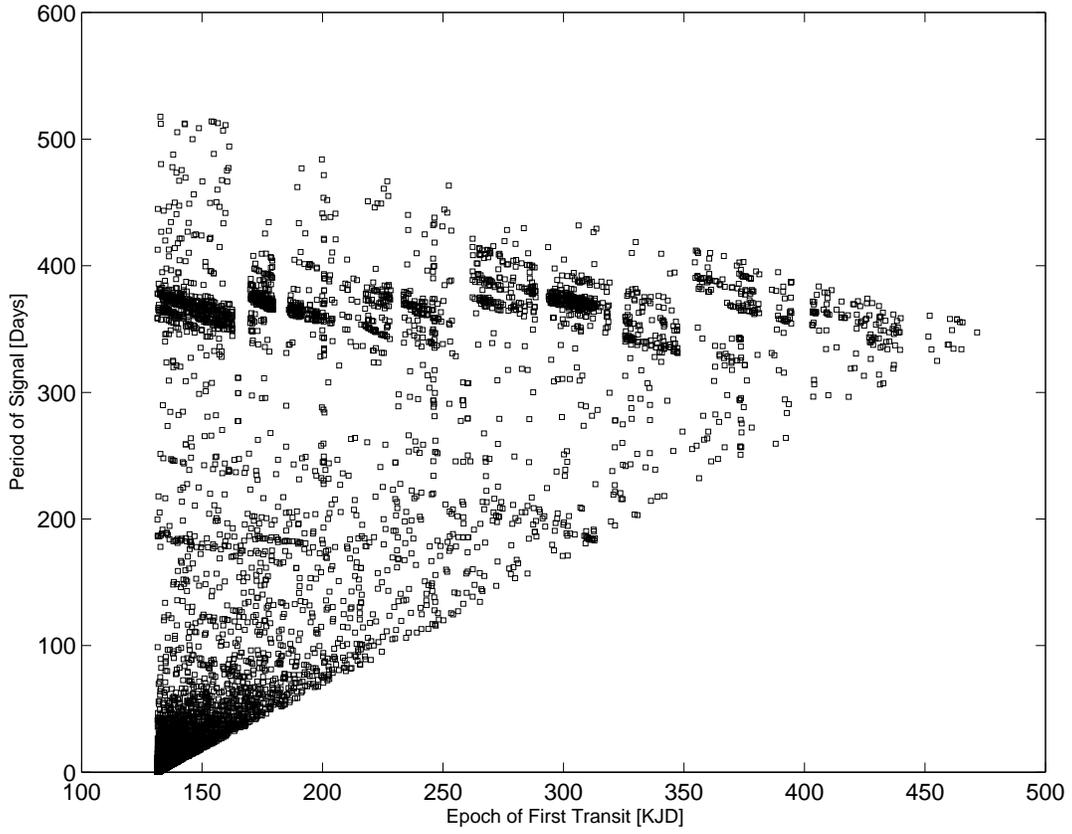}
\caption{Epoch and period of the 18,427 TCEs detected in the 12-quarter TPS run.  Periods are in days,
epochs are in Kepler-modified Julian Date (KJD), see text for definition.
\label{f3}}
\end{figure}

\begin{figure}
%\epsscale{.80}
\plotone{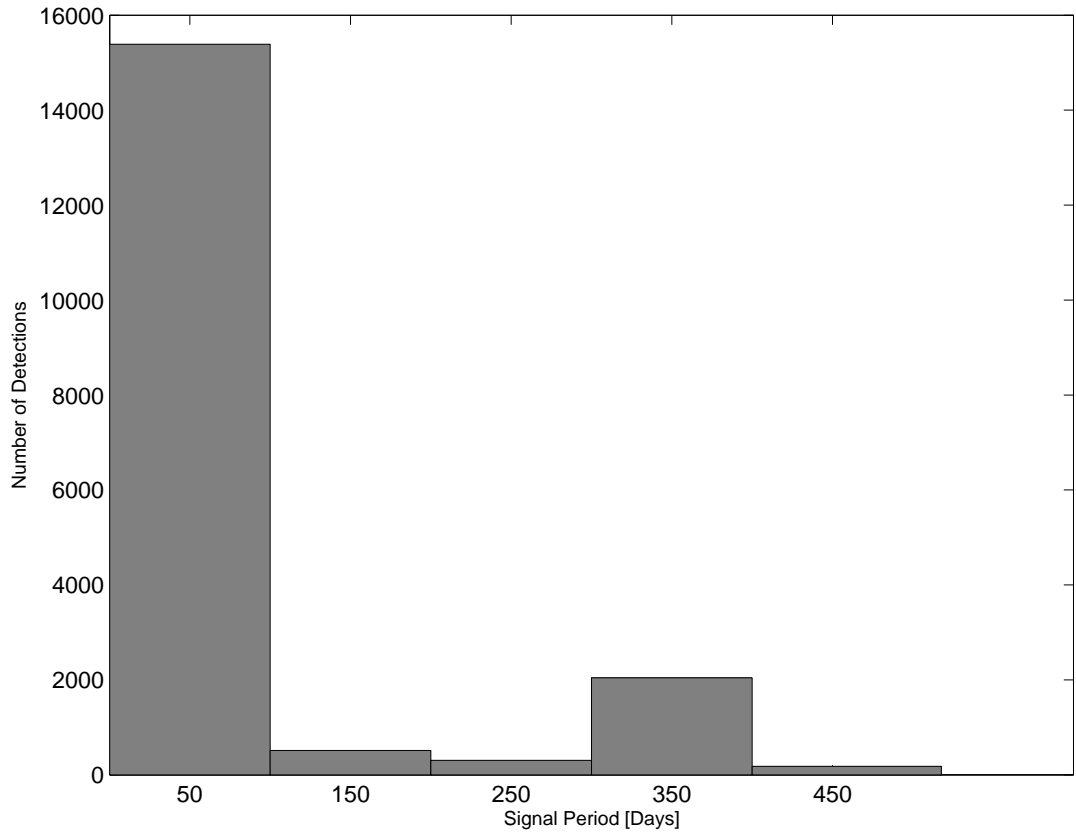}
\caption{Distribution of TCE periods.  The excess of detections at periods close to 1 year is due to the rotation
of a small number of image artifact channels about the focal plane as \kepler{} rotates about its axis.
\label{f4}}
\end{figure}

Figure \ref{f3} shows the epoch and period of the 18,406 detections, with period in days and epoch in Kepler-Modified Julian
Date (KJD), which is Julian Date - 2,454,833.0.  
While Figure \ref{f3} is relatively free of obvious artifacts,
there is an evident overabundance of detections at periods of approximately one year.  Figure \ref{f4} shows the distribution of
periods from Figure \ref{f3}; the overabundance is even clearer here, with 2,042 TCEs with periods between 300 and 400 days
as compared to 305 TCEs with periods of 200 to 300 days and 168 TCEs with periods of 400 to 500 days.  

\begin{figure}
%\epsscale{.80}
\plotone{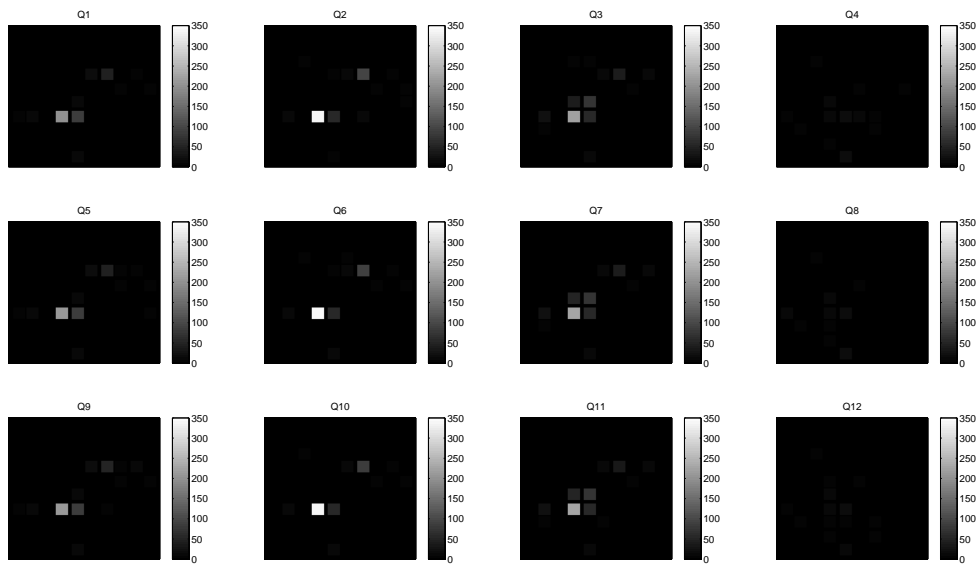}
\caption{Participation of \kepler{} output channels in TCEs with periods between 300 and 400 days.  The sub-plots
are all oriented such that modules 2, 3, and 4 are at the top.  Each column of sub-plots represents a common 
roll orientation of the spacecraft (i.e., quarters 1, 5, and 9 all correspond to the same orientation of the spacecraft).
The strongest
contributions come from Module 17, Output 2, which is known to exhibit temperature-dependent noise artifacts.  Other
strong contributors shown are Module 9, output 2; Module 13, Output 4; and Module 18, Output 2.  All of these 
channels are also known to exhibit unusually elevated noise, though not at the level of Module 17, Output 2.
Note that the pattern and intensity of misbehaving channels repeats annually, giving further evidence that the 
misbehaviors are driven by the spacecraft thermal environment, which itself repeats annually due to the quarterly
change in roll orientations.
\label{f5}}
\end{figure}

Figure \ref{f5} shows the participation of the various detector channels on the Kepler focal plane in TCEs with periods between 300 and 400
days:  each sub-image shows one quarter, and the relative intensity of each channel represents the participation, of that channel
in that quarter, in the 2,042 TCEs.  A small number of channels are disproportionately involved in these TCEs, mainly channels which
are known to suffer from excess noise due to issues in the readout electronics \citep{rolling-band,inst-performance}.  As \kepler{} rotates each 
quarter, certain stars will typically be imaged onto one of these misbehaving channels once per year; this will result in detections
on those stars with periods of approximately 1 year.  Efforts to manage the excess noise of these channels in \kepler{} data processing are ongoing.

\begin{figure}
%\epsscale{.80}
\plotone{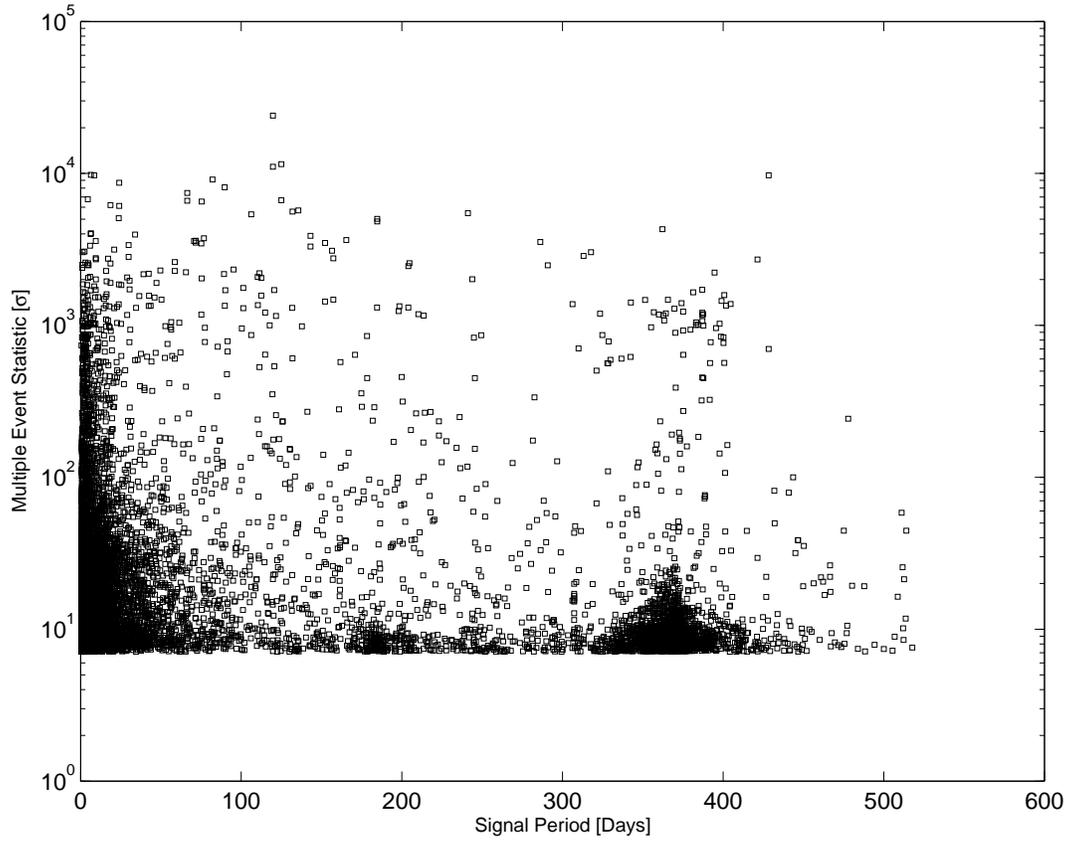}
\caption{Distribution of TCE periods and Multiple Event Statistics.
\label{f6}}
\end{figure}

\begin{figure}
%\epsscale{.80}
\plotone{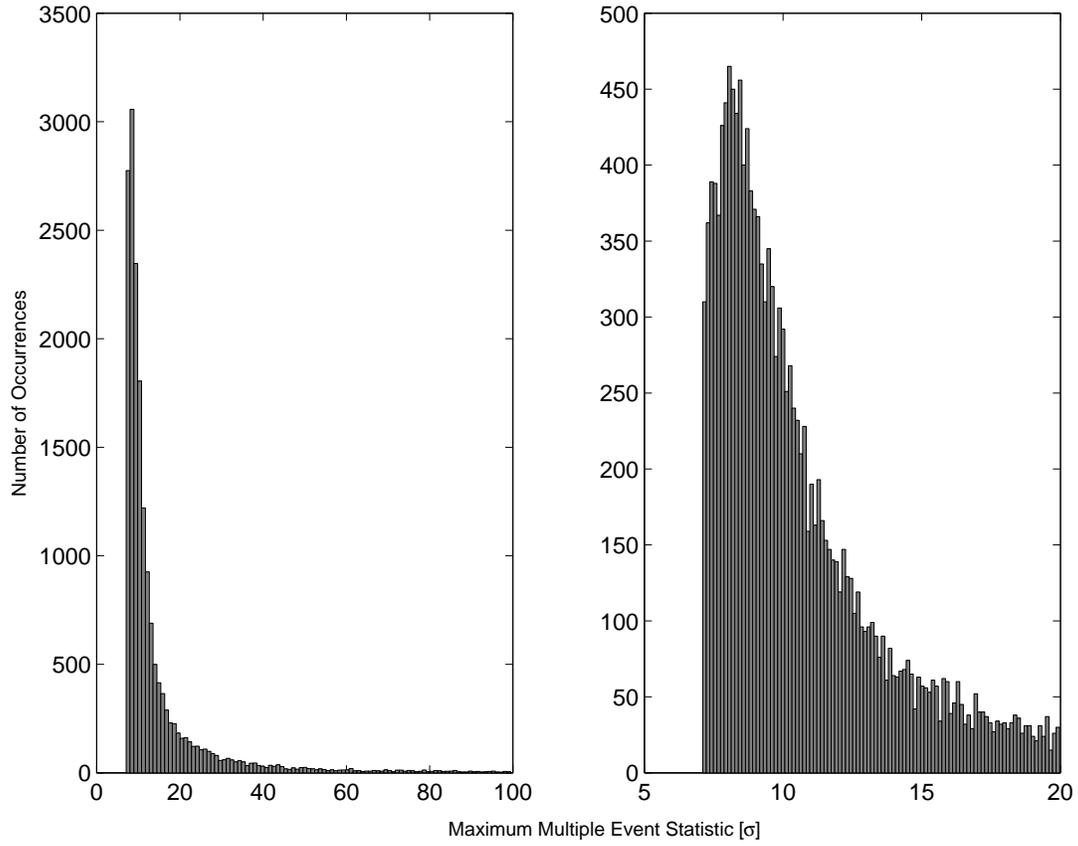}
\caption{Distribution of Multiple Event Statistics.  Left:  17,568 TCEs with Multiple Event Statistic of 100 or lower. 
Right:  15,018 TCEs with Multiple Event Statistic of 20 or lower.
\label{f7}}
\end{figure}

\begin{figure}
%\epsscale{.80}
\plotone{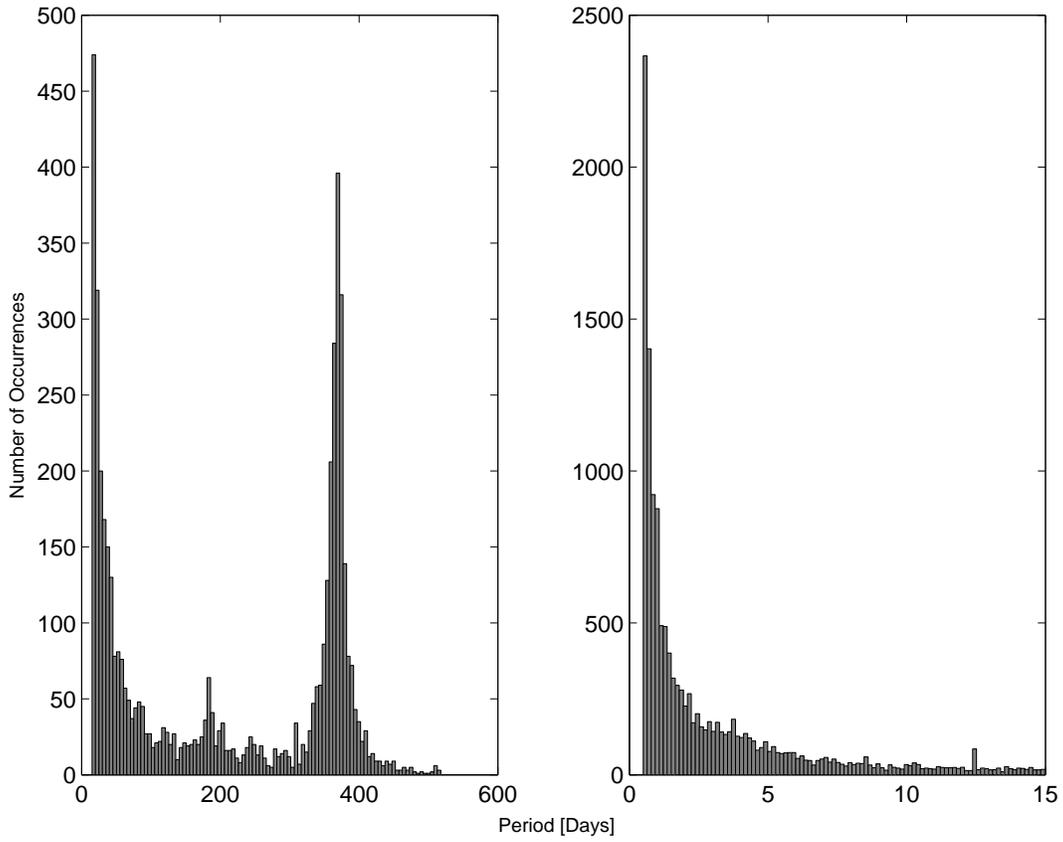}
\caption{Distribution of periods.  Left:  5,045 TCEs with periods greater than 15 days, with the data anomaly-driven excess
at approximately 1 year clearly visible.  Right:  13,382 TCEs with periods less than 15 days.
\label{f8}}
\end{figure}

Figure \ref{f6} shows the distribution of detections in the plane of orbital period and Multiple Event Statistic.  Note that the 
overabundance of detections at one year is completely dominated by relatively weak signals.  Figure \ref{f7}
shows the distribution of Multiple Event Statistics:  on the left is the distribution of 17,547 detections with Multiple Event Statistic
less than or equal to 100 $\sigma$; the right panel shows the same but for the 15,007 detections with Multiple Event Statistic
less than or equal to 20 $\sigma$.  Figure \ref{f8} shows the distribution of detected periods:  on the left is the 5,043
detections with periods over 15 days, on the
right is the 13,363 detections with  periods less than 15 days.  As compared to 
Figure 6 in \citet{pt2012}, the right side of Figure \ref{f8} is far more strongly peaked towards short periods.  Note that, in addition to
the excess of detections with periods close to 1 year, there is a smaller excess of detections with periods of 0.5 years.  This peak is
caused by the presence of two high-noise channels which are located symmetrically opposite one another on the focal plane, specifically
Module 17, Output 2, and Module 9, Output 2:  stars which are imaged onto one of these channels will be imaged onto the other 
6 months later.

\begin{figure}
%\epsscale{.80}
\plotone{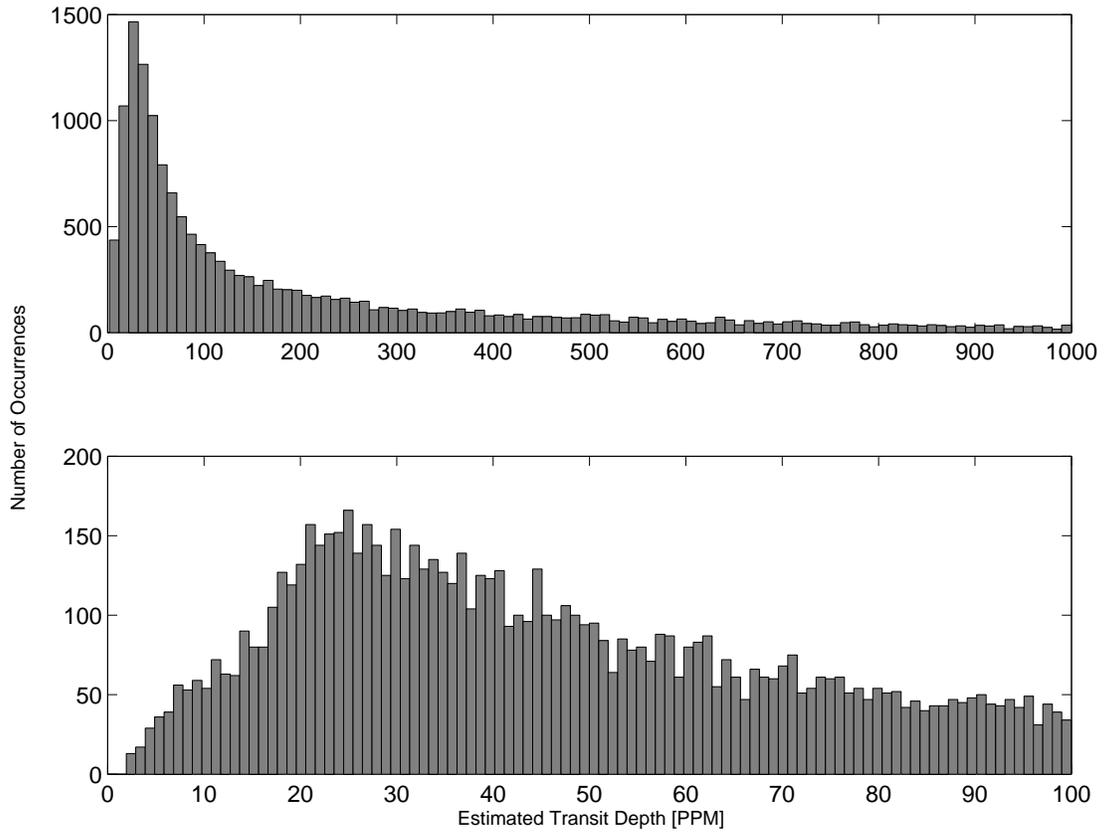}
\caption{Distribution of estimated transit depths.  Top:  16,115 signals with estimated depth of 1,000 parts per million (PPM) 
or less; bottom:  8,068 signals with 100 PPM or less.
\label{f9}}
\end{figure}

Figure \ref{f9} shows the distribution in estimated transit depths.  These depths are estimated from the event statistics and the
noise properties of each light curve, as described in \citet{pt2012}.  The top plot shows the 16,095 signals which have estimated
depths of 1,000 parts per million (PPM) or less; 
the bottom plot shows the the 8,060 cases with estimated depths of 100 PPM or less.  These sub-distributions contain 87.4\%
and 43.8\%, respectively, of all the detections in this dataset.  Comparing to the same transit depth ranges in \citet{pt2012},
we find that in the processing of the first 3 quarters of data the totals were 72.3\% and 13.2\%, respectively.  This increased sensitivity
to weaker transits is driven in the main by the vastly increased amount of data collected since the end of Quarter 3.

Figure \ref{duty-cycle} shows the distribution of transit duty cycles for all detections, where the transit duty cycle is defined to be the ratio of the
trial transit pulse duration to the detected period of the transit (effectively, the fraction of the time during which the TCE is in transit).  
The top plot shows all 18,406 TCEs, while
the bottom plot shows the 7,721 TCEs with transit duty cycle below 0.04.  Figure \ref{period-duty-cycle} shows the relationship between
period and transit duty cycle for all 18,406 detections.  As expected, the relationship is quantized due to the quantization of trial transit pulse durations 
utilized in the TPS detection algorithm, and as a consequence of this quantization the period and transit duty cycle are inversely proportional for
a given trial transit pulse duration. Figure \ref{period-duty-cycle}
also demonstrates why there is an abundance of events with transit duty cycles of approximately 0.002 shown in Figure \ref{duty-cycle}:
this is actually a reflection of the abundance of events with periods near 1 year, for which the possible transit duty cycles are all in the 
realm of $1\times10^{-4}$ to 0.002.

Detailed information on all TCEs which contributed to this analysis can be found at the NASA Exoplanet Archive:  
{\tt http://exoplanetarchive.ipac.caltech.edu/cgi-bin/ ExoTables/nph-exotbls?dataset=tce}.

\subsection{Comparison with Known Kepler Objects of Interest (KOIs)}

In order to gauge the performance of TPS as a detector of periodic transit-like phenomena, it is necessary to compare the
set of TCEs to a set of known events which can function as a ``ground truth''.  For this purpose, we use the list of Kepler Objects of Interest
(KOIs).  Out of the current set of KOIs (Burke, C.J. et al.2012, in preparation), we have selected 2,630 KOIs
which are judged reasonable for comparison to the TCE list:  these are KOIs for which the signal to noise ratio is high
enough to permit detection in TPS, the number of transits which fall within the 12 quarters of \kepler{} data is 3 or more, and
which do not fall on targets which were excluded from TPS processing.  The selected set of KOIs includes planet candidates,
known astrophysical false positives (mainly eclipsing binaries and background eclipsing binaries), and objects which have not
yet been characterized as planetary or non-planetary; for the purpose of the comparison, it is sufficient that each KOI be 
reasonably expected to produce a TCE.

The comparison of the KOI and TCE lists is complicated by the fact that any target star can have multiple KOIs and/or 
multiple TCEs, and the multiplicities of the two are obviously not guaranteed to agree.  As a first step, we compared the
number of TCEs on each KOI target star with the number of KOIs on those stars.  The result of this comparison is as follows:
\begin{itemize}
\item A total of 31 KOIs do not have a corresponding TCE
\item The remaining 2,599 KOIs were matched one-for-one by TCEs which occurred on the same target stars
\item 337 KOI target stars produced more TCEs than their known KOIs, resulting in a total
of 438 TCEs which fall on KOI targets but are not matched by known KOIs.
\end{itemize}

\subsubsection{Failure to Detect Short-period KOIs due to Data Artifacts}

Subsequent analysis of the KOIs which were not matched by TCEs showed that 21 out of the 31 had relatively short
periods, typically under 2 weeks.  Figure \ref{f10} shows the maximum Multiple Event Statistic as a function of period for a selected target
in this group.  The period and Multiple Event
Statistic of the KOI on this target star is indicated with a marker in the plot.  As shown in Figure \ref{f10}, the Multiple Event
Statistic is dramatically and systematically larger for long periods than for short periods, with a gross pattern of the 
Multiple Event Statistic rising as the square root of the period.  

The root cause of this pattern is a small number of strong transit-like data anomalies which are randomly distributed amongst
the flux time series.  During the folding process which results in Figure \ref{f10}, the anomalies are combined with background
noise to produce strong Multiple Event Statistics.  For short periods, the number of events folded together is large, thus there
are many background noise events combined with a single data anomaly; as a result, the Multiple Event statistic is relatively
small due to the dilution from the many background noise events.
For long periods, because the number of noise events is small, the data anomaly is
relatively undiluted and the resulting Multiple Event Statistic is relatively large.

A Multiple Event Statistic which is composed of one strong transit-like anomaly and multiple non-transit-like background noise
signals will not survive the Robust Statistic and chi-square vetoes, as it does not match the quantitative signatures of a true transit pulse
train which those vetoes require.  Unfortunately, as noted above, the ability of TPS to reject large numbers of such false detections
in a single light curve has been limited for reasons of computational performance:  the 1,000 combinations of period and transit
epoch which produce the strongest Multiple Event Statistics are searched, after which the search algorithm declares that no
transit signatures were found.  In the case of a target such as the one selected for Figure \ref{f10}, the 1,000 strongest signals
are all at the long-period end of the distribution, and the search iterations are exhausted before the actual signal at 3.766 days
is examined.  

In the limit where strong transit-like data anomalies are distributed uniformly and randomly throughout the dataset, there will
inevitably be some targets for which early quarters of data contain no anomalies but later quarters contain one or more.  If such
a target also contains a short-period, low-intensity transit signature, then the transit signature will be detectable only so long as
the data used for the detection was entirely acquired prior to the first anomaly occurrence.  This appears to be the case for the 21 instances
of short-period, low-intensity KOIs which were not detected by the most recent TPS run.  Note that this is one of those unusual 
situations in which a 12 quarter dataset does not permit detection of a signal which was apparent in a 3- or 6-quarter dataset.

\subsubsection{Matching of KOI and TCE Ephemerides}

Detection of a TCE on a KOI target is a necessary but not sufficient condition to determine that the TCE is a detection of the KOI.  An additional
requirement is that the TCE and KOI are referring to the same transit signature.  This is typically best determined by matching the ephemerides
of the two signatures.  For this purpose we use an ephemeris-matching calculation described in Appendix \ref{matching}.  The
resulting match parameter varies from a value of zero, indicating no match whatsoever, to a value of one, indicating a perfect
match within the limits of the \kepler{} data and data processing algorithm.  In the case of a target star which has multiple
KOIs and/or multiple TCEs, it is necessary to attempt to correctly match each KOI with the corresponding TCE.  A subtlety in
this process is that it is at least conceivable that multiple KOIs will be best matched by the same TCE.  For example, consider
a target which has two KOIs, with periods of 0.5 and 1.0 years, and three TCEs, with periods of 0.5, 0.1, and 0.03144 years.  Depending
on the detailed transit timings, it is at least conceivable that the TCE with the 0.5 year period will be the best match out of the 3 TCEs
for both the 0.5 year and 1.0 year period KOIs.  In order to ensure that each TCE is paired with one and only one KOI, the following
approach is used:
\begin{itemize}
\item Compute the ephemeris matches for all $n_{\rm KOI} \times n_{\rm TCE}$ possible matches between KOI and TCE
\item Find the best match in that matrix, and pair the corresponding KOI and TCE with one another
\item Eliminate both the KOI and the TCE which have now been paired
\item Repeat the exercise with the remaining $(n_{\rm KOI}-1) \times (n_{\rm TCE}-1)$ possible matches, and iterate until
either the number of TCEs or the number of KOIs on the given target star are exhausted.
\end{itemize}

Figure \ref{f11} shows the value of the ephemeris match between each of the 2,599 KOIs and the TCE on that star
which provided the closest match.  The values in Figure \ref{f11} are sorted into descending order.  Of the 2,599 match values,
only 104 are less than 1.0, with 2,495 identically equal to 1.  Of these 104 cases, 91 are either harmonic mismatches
between the TCE and the KOI (especially in cases where the KOI period is under the 0.5 day minimum period used in TPS) or cases
in which the KOI timing was determined using only data from early quarters, resulting in errors when extrapolating the timing
to the full 12 quarters used in this analysis.  The remaining classes of discrepancy between TCE and KOI are as follows:
\begin{itemize}
\item In 8 cases, transit timing variations (TTV) cause confusion for TPS, which is explicitly designed to find periodic transit signatures; this generally
results in a tremendous period mismatch between the KOI timing and the TCE, since TPS will usually detect a tiny subset of all transits.
\item In 3 cases, the KOI and the TCE have inconsistent transit timing signatures, but both signatures appear valid.  In each of these cases it is
assumed that TPS has identified a heretofore-unknown transit signature on the KOI target, but then failed to detect the known KOI during
the multiple-planet search which followed detection of the new TCE.  For this reason, these cases are classified as failures of the TPS
algorithm to recover the known KOIs.
\item In 2 cases the KOI timing clearly produces a transit signature and the TCE timing clearly does not.
\end{itemize}

\subsubsection{Conclusion of TCE-KOI Comparison}

Out of 2,630 KOIs which could be expected to produce TCEs,  44 did not produce TCEs.  This includes 31 cases in which there
was no TCE and 13 cases in which a TCE was produced but the timing of the TCE did not match the timing of the KOI, even when
``near misses'' such as harmonic or sub-harmonic detection are taken into account.  This yields a KOI recovery rate of 2,586 out of
2,630, or 98.3\%.

\subsubsection{Transit Duty Cycle of TCEs Matched to KOIs}

\begin{figure}
%\epsscale{.80}
%\plotone{duty-cycle.eps}
\plotone{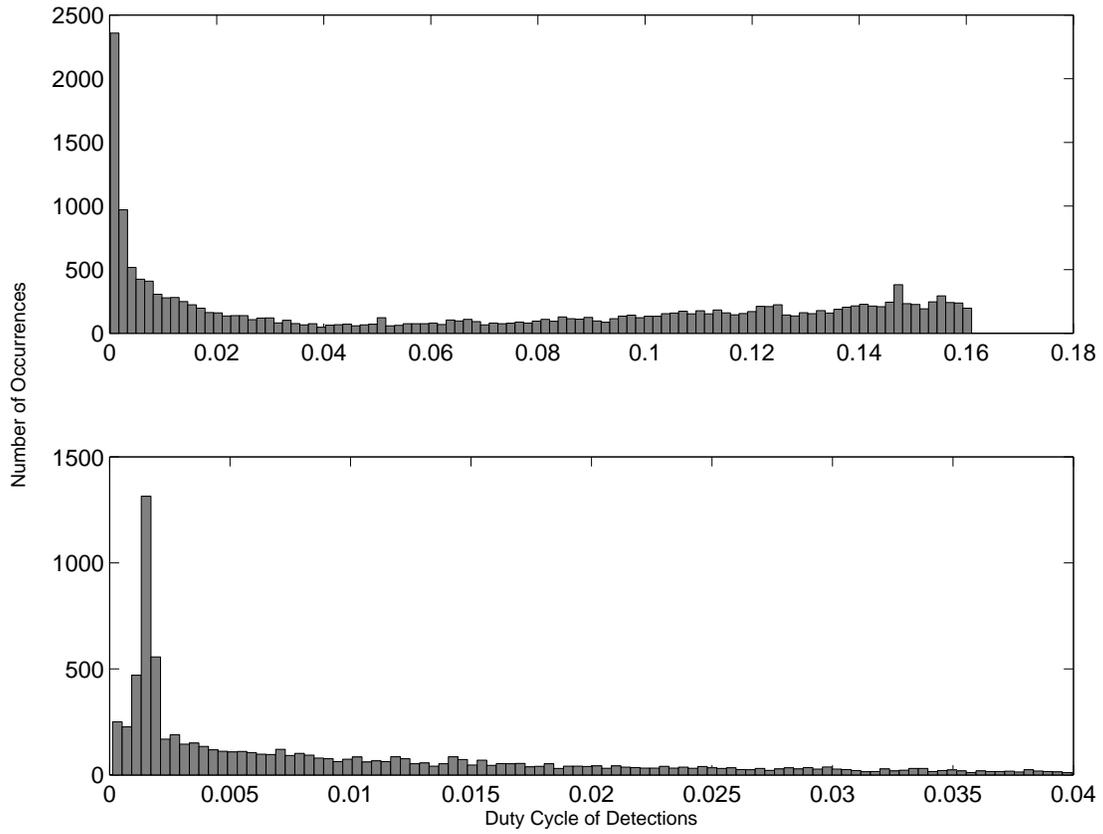}
\caption{Distribution of transit duty cycles.  Top:  all TCEs.  Bottom:  7,729 TCEs with transit duty cycle below 0.04.
\label{duty-cycle}}
\end{figure}

\begin{figure}
%\epsscale{.80}
%\plotone{period-duty-cycle.eps}
\plotone{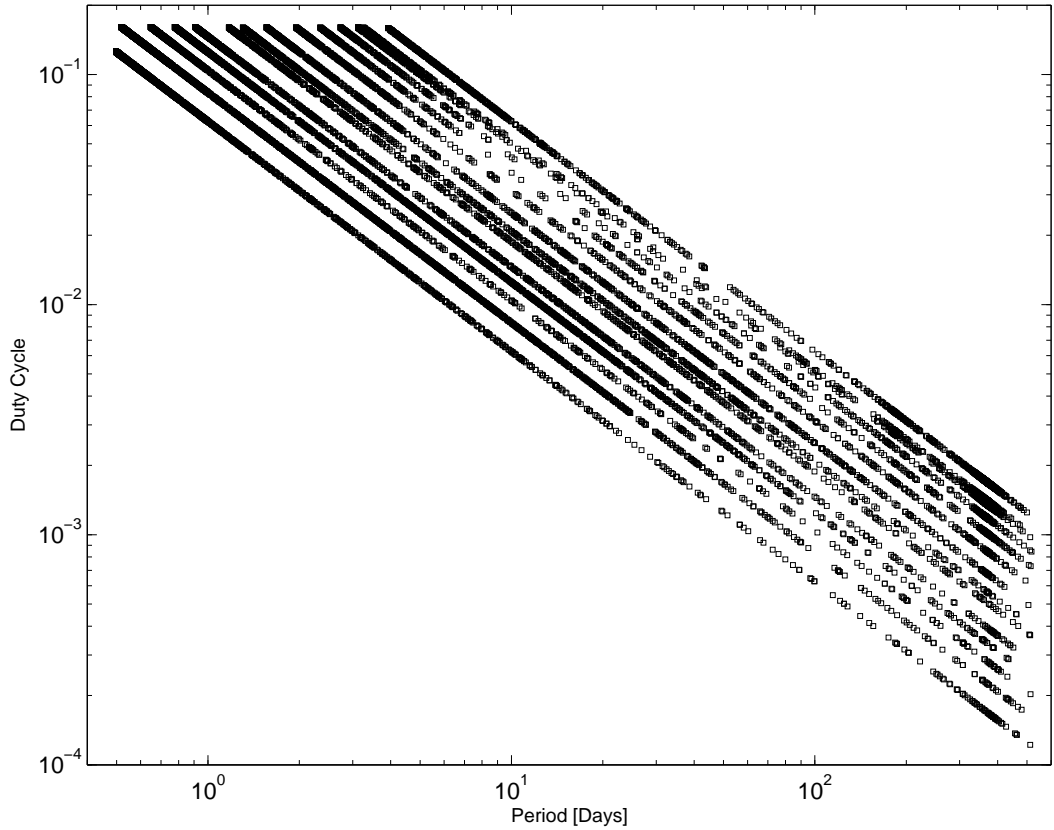}
\caption{Relationship between period and transit duty cycle for all TCEs.  The structure observed is driven by the fact that TPS
uses a small number of fixed trial transit pulse durations for its searches, and by the fact that at a given trial transit pulse duration
the transit duty cycle is inversely proportional to the TCE period.
\label{period-duty-cycle}}
\end{figure}

\begin{figure}
%\epsscale{.80}
%\plotone{f10.eps}
\plotone{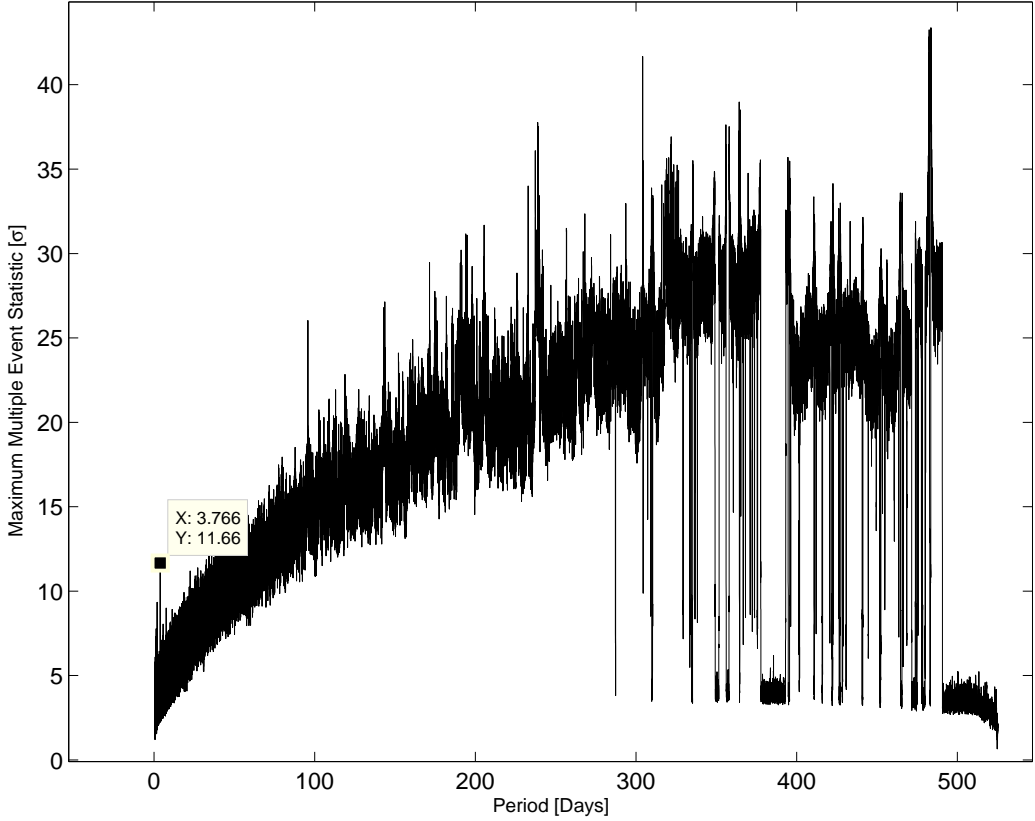}
\caption{Maximum Multiple Event Statistic as a function of period for a sample target.  In this target, the KOI period of 3.766
days is shown at the marker, with a Multiple Event Statistic of 11.66 $\sigma$.  One or more artifacts in the flux time series are
causing the large number of larger Multiple Event Statistic values at longer periods.  Because of the 1,000 iteration limit on
rejecting strong signals and re-searching for better but weaker signals, this KOI is not detected:  the 1,000 iterations are exhausted
before all of the false alarms in the figure can be rejected.
\label{f10}}
\end{figure}

\begin{figure}
%\epsscale{.80}
%\plotone{f11.eps}
\plotone{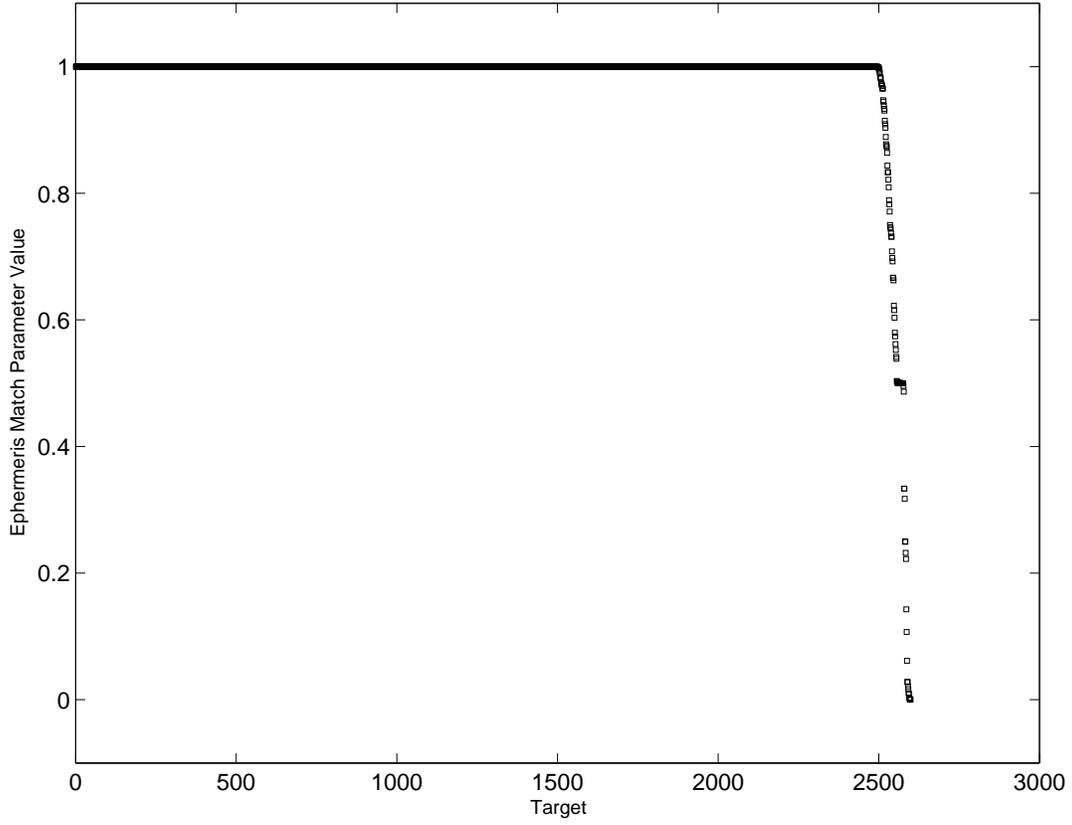}
\caption{Value of the ephemeris-match parameter described in the text across all 2,608 TCEs which are matched to known KOIs.
Only 113 of the values are not identically equal to 1.
\label{f11}}
\end{figure}

\begin{figure}
%\epsscale{.80}
%\plotone{koi-tce-duty-cycle.eps}
\plotone{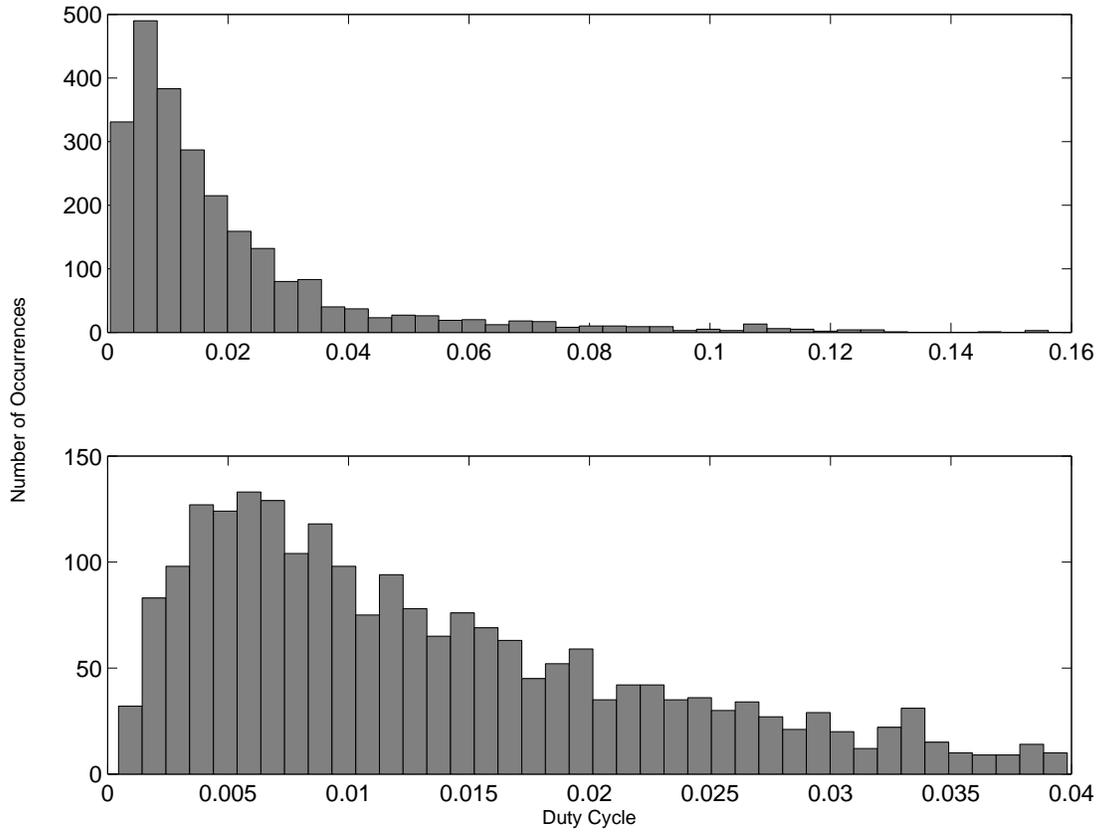}
\caption{Distribution of transit duty cycles for TCEs successfully matched with KOIs.  Top:  2,495 cases in which the ephemeris match is 
identically equal to 1.  Bottom:  2,205 cases in which the ephemeris match is identically equal to 1 and the transit duty cycle is less than
0.04.
\label{koi-tce-duty-cycle}}
\end{figure}

Figure \ref{koi-tce-duty-cycle} shows the distribution of TCE transit duty cycles for the 2,495 cases in which the TCE-KOI ephemeris match
is identically equal to 1, as well as the distribution for the 2,205 cases in which the ephemeris match is identically equal to 1 and the
transit duty cycle is below 0.04.  When compared to Figure \ref{duty-cycle}, which shows the transit duty cycle for all TCEs, two differences are
instantly apparent.  First, and least surprisingly, the spike in transit duty cycle values around 0.002 which is visible in Figure \ref{duty-cycle}
is absent from Figure \ref{koi-tce-duty-cycle}.  This is because the spike in the former is due to the spurious, anomaly-driven detections
at 1 year period which are caused by CCD readouts with unusually strong noise properties; these spurious detections are not present
in the set of KOIs, thanks to the greater degree of scrutiny on KOIs which allows elimination of such false detections.  Second, the KOI
transit duty cycle distribution shows a monotonic reduction in the number of KOIs as the transit duty cycle is increased; the TCE distribution shows
a reduction from 0.01 to 0.04 transit duty cycle, and an increase from 0.04 to 0.16.  Quantitatively, while 58\% of  all TCE detections in 
Figure \ref{duty-cycle} have a transit duty cycle of 0.04 or greater, only 12\% of all KOIs in Figure \ref{koi-tce-duty-cycle} have transit duty cycle 
above 0.04.  The implication is that the long transit duty cycle TCEs are most likely dominated by false positive detections, and that further
reduction in the maximum allowed transit duty cycle from the current value of 0.16 would result in further reduction of the fraction of false positive TCEs, though of course
some study would be needed to determine an optimum threshold for the transit duty cycle.

\section{Conclusions}

The \kepler{} Transiting Planet Search (TPS) algorithm has been run on 192,313 targets in the \kepler{} field of view, 
including 112,321 targets which have been observed near-continuously for the first 12 quarters of the mission.  Potential
signals of transiting planets were detected on 11,087 of these targets.  When subjected to further searches for 
multiple planets, the total number of detected signals grew to 18,406.  Comparison with a known and vetted set of transit-like astrophysical signatures,
the Kepler Objects of Interest (KOIs), demonstrates that within the parameter regime of the search algorithm and the KOIs the recovery
rate of known events is 98.3\%.  

\section{Acknowledgements}

Funding for this mission is provided by NASA's Space Mission Directorate.  The
contributions of Hema Chandrasekaran and Chris Henze have been essential in the studies documented
here.
 \appendix

 \section{Construction of the Robust Statistic Veto}\label{robustStat}

The first step in constructing the Robust Statistic is to generate the transit model pulse train.  This consists of a train of square wave pulses that are positioned 
at the locations of the transits as determined by the period and epoch associated with the Multiple Event Statistic.  Let $\mathbf{s}$ be this model pulse train 
vector.  The pulse train $\mathbf{s}$ and the data, or flux time series, $\mathbf{x}$ are each whitened to eliminate the effect of stellar variations.  The 
whitened model and data vectors, $\tilde{\mathbf{s}}$ and $\tilde{\mathbf{x}}$ (where `$\sim$' denotes a whitened vector), are then windowed to remove 
out-of-transit samples.  The resulting whitened, windowed, transit model $\tilde{\mathbf{s}}$ is then robustly fit to the whitened, windowed, 
data $\tilde{\mathbf{x}}$ to generate a diagonal matrix of fit weights $\mathbf{W}$.  The Robust Statistic (RS) is then calculated as:
\begin{equation}\label{rsdef}
{\rm RS} = \frac{ \tilde{\mathbf{s}}^T \mathbf{W} \tilde{\mathbf{x}} }{\sqrt{\tilde{\mathbf{s}}^T \mathbf{W} \tilde{\mathbf{s}} }} ,
\end{equation}
where $T$ denotes the transpose of a vector, and where Equation \ref{rsdef} is applied only to data samples within the transit windows described above.

In the limit in which the data vector $\mathbf{x}$ and the model vector $\mathbf{s}$ are well-matched in shape and duration, the matrix $\mathbf{W}$ will 
approach the identity matrix and the RS as defined in Equation \ref{rsdef} will be approximately equal to the Multiple Event Statistic.  In reality, the match between
data and model is imperfect:  the transits in the model vector are represented as square wave pulses rather than true transit shapes, and in general the 
duration of the trial transit pulse and the true transits will not be identically matched to one another.  In studies of known transiting planet systems, this 
mismatch can lower the RS by about 10\% compared to the Multiple Event Statistic.

Now consider a situation in which the Multiple Event Statistic is constructed from folding a single, extremely strong transit-like signature over two or more
events which are consistent with statistical fluctuations, which is a typical case of non-uniform-depth events being combined into a Multiple Event Statistic
which lies above threshold.  Because the fit is performed robustly, the weak transit-like signatures will ``out-vote'' the strong one, leading to near-unity weights
for the weak events and near-zero weights for the strong event.  When the weights in this instance are combined with the data and model vectors as shown
in Equation \ref{rsdef}, the result will be a low value for RS.  It is in this way that the RS permits events with significant transit depth mismatches to be vetoed while
preserving events with relatively uniform transit depths.
 
 \section{Threshold Crossing Event Vetoes using Chi-Square Discriminators}\label{chisquare}
 
The basic idea behind the construction of the test statistic is to break up the detection statistic into several contributions and compare each 
observed contribution with what is expected (\cite{Allen:2004gu}).  Note that with some basic assumptions on the detector noise (namely, that the 
noise after whitening is zero mean, unit variance, and uncorrelated) the expectation values of test statistics formulated below are independent of whether or
not a signal is present in the data, making them ideal discriminators for noise events.

Beginning with the single event statistic time series $z(n)$:
\begin{equation}
\label{SES}
\nonumber
z(n)  =  \frac{\mathbb{N}(n)}{\sqrt{\mathbb{D}(n)}},
\end{equation}
where
\begin{eqnarray}\label{define-ND}
\mathbb{N}(n) & \equiv & \sum_{i=1}^{M} 2^{-\min(i,M-1)}\left\{\left[\frac{x_i}{\hat{\sigma^2_i}}\right]*\tilde{s}_i\right\}(n) 
=  \sum_{i=1}^{M} \mathbb{N}_i(n), \\ \nonumber
\mathbb{D}(n) & \equiv & \sum_{i=1}^{M} 2^{-\min(i,M-1)} \left[\hat{\sigma}_i^{-2} * \tilde{s}_i^2\right](n)
=  \sum_{i=1}^{M} \mathbb{D}_i(n),
\end{eqnarray}
where $M$, $x_i$, $\hat{\sigma}_i$, and $\tilde{s}_i$ are defined in Appendix A of \citet{pt2012}.  
Qualitatively, the time series $\mathbb{N}(n)$ represents the amplitude of a transit-like signal centered at sample $n$, $\mathbb{D}(n)$ represents
the square of the noise limit for detecting a transit-like signature at sample $n$; $z(n)$ therefore represents the significance of a transit-like signature
detected at sample $n$.  
Equation \ref{define-ND} also defines quantities 
$\mathbb{N}_i$ and $\mathbb{D}_i$:  these are the contributions to $\mathbb{N}$ and $\mathbb{D}$, respectively, from frequency band $i$.  
Choosing a particular point in transit duration, period, and epoch space, $\{D,T,t_0\}$, selects out a set of data samples $\{A\}$,
one for each transit,
that start with the sample corresponding to the epoch $t_0$ and are spaced $T$ samples apart.  These samples form a subset of $\{n\}$, $A \subset \{1,2,...,P\}$, where $P$ is the number of transits in the dataset.  
The Multiple Event Statistic is then constructed as:
\begin{equation}
\label{MES}
Z(D,T,t_0) = \sum_{i \in A} \mathbb{N}(i) \slash \sqrt{ \sum_{i \in A} \mathbb{D}(i) } \ \ .
\end{equation}

One version of the $\chi^2$ can be constructed by focusing on the wavelet contributions to the Single Event Statistics.  
If we start now with (\ref{SES}), we can make the identifications:
\begin{equation}
z_i(n)  =  \frac{\mathbb{N}_i(n)}{\sqrt{\mathbb{D}(n)}} \\
\end{equation}
\begin{equation}
q_i(n)  =  \frac{\mathbb{D}_i(n)}{\mathbb{D}(n)} \\ ,
\end{equation}
where now the $z_i(n)$ are the actual contributions the the SES time series from the $i$'th wavelet component and $q_i(n)$ are the corresponding expected contributions.  Now the $\chi^2$ statistic can be formed:
\begin{equation}
\Delta z_i(n) = z_i(n) - q_i(n)z(n)
\end{equation}
\begin{equation}
\chi^2(n) = \sum_{i=1}^M \frac{[\Delta z_i(n) ]^2}{q_i(n)} \ \ .
\end{equation}
Using the previously mentioned noise assumptions, this statistic should be $\chi^2$ distributed with $M-1$ degrees of freedom; due to leakage between the wavelet components it turns out to be
gamma distributed in actual practice.  
We have a value for this statistic at each $n$, so we can form a coherent statistic by adding up the points that contribute to the 
Multiple Event Statistic at times $j$ where $j \in A$.  This will give us, $\chi_{(1)}^2$,
\begin{eqnarray}
\nonumber
\chi_{(1)}^2 & = & \sum_{j \in A} \chi^2(j) \\
\nonumber
           & = &  \sum_{j \in A} \sum_{i=1}^M \frac{[\Delta z_i(j)]^2}{q_i(j)} \\
           & = & \sum_{j \in A} \sum_{i=1}^M \frac{\Delta z_{ij}^2}{q_{ij}} \ \ ,
\end{eqnarray}
where the $\Delta z_{ij}$ and $q_{ij}$ have been introduced for notational convenience.  Using the previous assumptions on noise and assuming a perfect match between the signal and template, this statistic is $\chi^2$-distributed with $P(M-1)$ degrees of freedom.

Another version of the $\chi^2$ statistics can be constructed by examining the $P$ temporal contributions to the Multiple Event Statistic.  
To begin, Equation \ref{MES} can be rewritten using the quantities defined for notational convenience:
\begin{eqnarray}
\nonumber
Z & = & \frac{\sum_{j \in A} \mathbb{N}(j)}{\sqrt{ \sum_{j \in A} \mathbb{D}(j) }} \\
\nonumber
  & = & \frac{\sum_{j \in A} \sum_{i=1}^M \mathbb{N}_{i}(j)}{\sqrt{\sum_{j \in A} \sum_{i=1}^M \mathbb{D}_{i}(j)}} \\
  & = & \frac{\sum_{j \in A} \sum_{i=1}^M \mathbb{N}_{ij}}{\sqrt{\sum_{j \in A} \sum_{i=1}^M \mathbb{D}_{ij}}} \ \ .
\end{eqnarray}
 Now, choosing to examine the contributions to the Multiple Event Statistic from each $j \in A$, 
\begin{equation}
Z_j = \frac{\sum_{i=1}^M \mathbb{N}_{ij}}{\sqrt{\sum_{k \in A} \sum_{i=1}^M \mathbb{D}_{ik}}}
\end{equation}
\begin{equation}
Q_j = \frac{\sum_{i=1}^M \mathbb{D}_{ij}}{\sum_{k \in A} \sum_{i=1}^M \mathbb{D}_{ik}} \ \ ,
\end{equation}
where now $Z_j$ are the actual temporal contributions to the Multiple Event Statisic and the $Q_j$ are the 
expected contributions.  Now, $\chi_{(2)}^2$ can be constructed:
\begin{equation}
\Delta Z_j = Z_j - Q_j Z
\end{equation}
\begin{equation}
\chi_{(2)}^2 = \sum_{j \in A} \frac{\Delta Z_j^2}{Q_j} \ \ .
\end{equation}
Under the previous noise assumptions, this statistic is $\chi^2$-distributed with $P-1$ degrees of freedom. Since we have summed over the wavelet contributions
prior to computing this statistic it avoids the leakage issue and turns out to be a much more powerful discriminator.  Note that dozens of other version of the chi-square veto have been
formulated and investigated with real data, and indeed an infinity of such statistics exists.  These two versions give us the greatest detection efficiency while simultaneously minimizing the false alarm rate.

The results quoted in what follows are subject to a subtle issue discovered after the Q1-Q12 run was completed.  The whitening coefficients in the calculation should be robust against the presence of a signal in the data since they are computed using a moving circular median absolute deviation.  However, the $\chi^2$ statistics are very sensitive to any signal dependence of the whitening coefficients, however small it may be, due to the way in which they are constructed.  The code is now being re-written so that in-transit cadences are first gapped and filled to re-compute the whitening coefficients for use in the $\chi^2$ calculation.  This should explicitly remove the signal dependence and give us more vetoing power.   

Based on analysis of known true-positive and expected false-positive targets, TPS uses the following discriminators in vetoing
false-positive detections:%
\begin{eqnarray}
X_{(1)} \equiv \frac{Z\sqrt{P(M-1)}}{\sqrt{\chi^2_{(1)}}}, \\ \nonumber
X_{(2)} \equiv \frac{Z\sqrt{P-1}}{\sqrt{\chi^2_{(2)}}}. \\ \nonumber
 \end{eqnarray}
 In words, the Multiple Event Statistic for a possible detection is divided by the square-root of the reduced chi-square for each of the
 chi-square statistics computed above, resulting in two discriminators.  
 
 \section{Ephemeris-Matching Calculation Used in KOI-TCE Comparisons}\label{matching}
 
 Consider a TCE which is characterized by its period $T_{\rm TCE}$, epoch $t_{\rm TCE}$, and trial transit pulse duration $D$; on 
 the same target star, consider a KOI which is characterized by its period $T_{\rm KOI}$ and epoch $t_{\rm KOI}$.  The following
 calculation can be used to determine whether the two ephemerides represent a good match or a poor match in transit timing.
 
 First, of the two periods, define $T_{\rm short}$ to be the shorter, and $t_{\rm short}$ to be the corresponding epoch (i.e., if the
 KOI has a shorter period, then $T_{\rm short}\equiv T_{\rm KOI}$ and $t_{\rm short}\equiv t_{\rm KOI}$); define $T_{\rm long}$
 and $t_{\rm long}$ to be the period and epoch of the ephemeris with the longer period.  The ephemeris matching parameter
 is the fraction of transits predicted by $(T_{\rm short},t_{\rm short})$ which fall within $D/2$ of one of the transits predicted by
 $(T_{\rm long},t_{\rm long})$.  
 
 The reason for using the fraction of short-period transits which are predicted is that there will always be more short-period transits
 than long-period ones.  In the case of an extremely large mismatch in periods between the two ephemerides (for example, a 3 day
 and a 300 day period), it is possible for all of the longer-period transits to fall close to transits of the shorter period, but the reverse
 is not true.  Thus, in cases of extreme mismatch in period, using the fraction of short-period transits as the metric ensures that
 matching parameter has a low value, whereas the fraction of long-period transits which fall near a short-period transit can be
 large, and thus use of the long-period transits in this way could result in a large value of the matching parameter even though the
 ephemerides are wildly mismatched.
 
The duration of the trial transit pulse must be included because the finite pulse width and the finite duration of a real transit result
in a family of nearly-degenerate (period,epoch) combinations.  For example, a dataset which contains 3 transits of 13 hour duration
at 365 day period would be well-matched by a model transit with 365 day period, but almost equally well by a transit with 364.9 day 
period or 365.1 day period.  The matching parameter takes this degeneracy into account by requiring that the short-period transits
be within one-half of a trial transit duration of the long-period transits.  The duration ``smearing'' is applied to the longer-period
ephemeris because, in a case with a huge period mismatch, applying it to the short-period ephemeris could result in duty-cycle
problems.  For example, consider the match between a 365 day period ephemeris with 13 hour duration and a 1 day period 
ephemeris.  Applying the pulse duration smearing to the short-period ephemeris would result in a duty cycle greater than 0.5;
applying the smearing to the long-period ephemeris ensures that such absurd combinations of parameters do not occur.
%==================================================

\clearpage

\clearpage

\clearpage

\clearpage

\clearpage

\clearpage

\clearpage

\clearpage

\clearpage

\clearpage

\clearpage

\clearpage

\clearpage

\clearpage

\clearpage

\end{document}